# Comparison of the superconducting critical transition temperature of LaBaCaCu$_3$O$_{7-d}$ and NdBaCaCu$_3$O$_{7-d}$

V.P.S. Awana[*], M.A. Ansari, Rashmi Nigam[$], Anurag Gupta, R.B. Saxena, and H. Kishan

National Physical Laboratory, K.S. Krishnan Marg, New Delhi 110012, India.

LaBaCaCu$_3$O$_7$ (La:1113) and NdBaCaCu$_3$O$_7$ (Nd:1113) compounds are synthesized by solid state reaction route. Both compounds crystallize in tetragonal structure. Four probe resistivity measurements showed that superconducting transition temperature ($T_c$) is 73 K and 41 K respectively for La:1113 and Nd:1113 compounds. Considering the fact that for RE:123 (REBa$_2$Cu$_3$O$_7$) compounds the $T_c$ hardly depends upon the choice of RE (Rare earth, except Ce, Pr and Tb), the current nearly two fold increase in $T_c$ of La:1113 when compared to Nd:1113, warrants serious discussion.

## I. INTRODUCTION

There are several RE (rare earth)-based HTSC (High Temperature Superconductor) families of which more extensively studied are REBa$_2$Cu$_3$O$_7$ (RE:123), REBa$_2$Cu$_4$O$_8$ (RE:124) and RE$_2$Ba$_4$Cu$_7$O$_{14}$ (RE:247) [1,2]. RE-based 123, 124 and 247 superconductors are known for all RE except Ce, Pr and Tb. While Ce and Tb do not form the required structure, the case of Pr is particularly interesting in that although the structure is formed the material is insulating, which is generally attributed to its magnetic interaction with neighboring Cu-O conduction band [3,4]. Another relatively less explored, RE-based family is RE:1113 (REBaCaCu$_3$O$_7$), which in many ways is unique in terms of phase formation and crystal structure. Unlike RE:123 and 124, RE:1113 compounds form only with lighter rare earths such as La, Pr and Nd [5]. In present study we show results of phase formation and superconductivity in La:1113 and Nd:1113 compounds, and discuss that why the $T_c$ of the former is nearly two times that of the latter.

## II EXPERIMENTATION

LaBaCaCu$_3$O$_7$ (La:1113) and NdBaCaCu$_3$O$_7$ (Nd:1113) compounds are synthesized by solid state reaction route using ingredients $La_2O_3$, $CaCO_3$, $BaCO_3$ and CuO. Calcinations were carried out on mixed powders at 900 $^0$C, 925 $^0$C, 950 $^0$C and 960 $^0$C each for 24 hours with intermediate grindings. These samples were further annealed in a flow of oxygen at 450 $^0$C for 40 hours and subsequently cooled slowly, over a span of 20 hours, to room temperature. X-ray diffraction (XRD) data were obtained at room temperature (MAC Science: MXP18VAHF[22]; Cu$K_\alpha$ radiation). Resistivity measurements in the temperature range of 12 - 300 K, were performed using a four- probe method in a close cycle refrigerator.

## III RESULTS AND DISCUSSION

The X-ray diffraction (XRD) of both the samples studied are phase pure in nature (XRD not shown). Lattice parameter determined from the XRD data for both La:1113 and Nd:1113 samples revealed that these samples crystallize in a tetragonal structure. The lattice parameters are: $a = 3.869(7)$ Å, and $c = 11.645(6)$ Å, for La:1113 and $a = 3.862(6)$ Å, and $c = 11.643(3)$ Å, for Nd:1113 sample. These compounds do crystallize in tetragonal structure despite having oxygen content close to 7.0 due to intermixing of Ca, Ba and RE lattice cites in normal RE:123 structure, for schematic cell see Fig. 1. Fig. 2 shows the R(T) plots for both La:1113 and Nd:1113 compounds in temperature range of 12 to 300 K. We observe that La:1113 has a $T_c$ = 71 K that is much higher than that for Nd:1113 $T_c$ = 42 K. Note also that these $T_c$s are typically much less than $T_c$ = 90 K seen in RE:123 system.

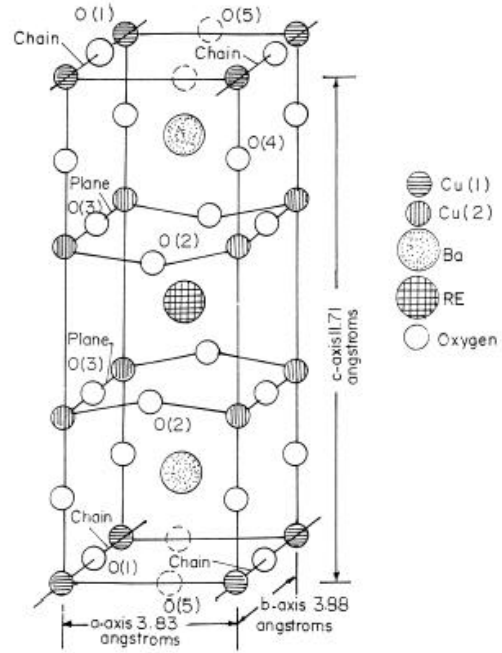

Fig.1 Schematic unit cell of $REBa_2Cu_3O_7$.

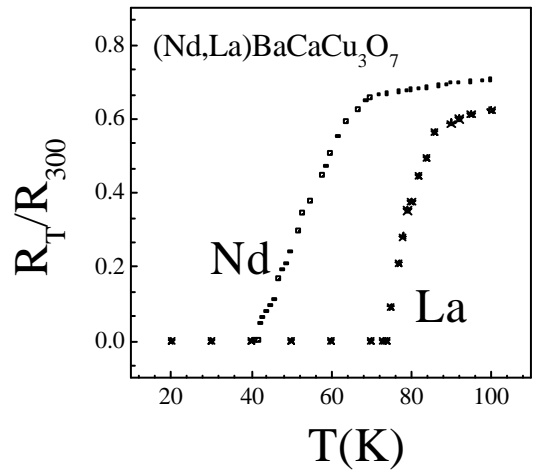

Fig. 2 Resistance versus temperature plots for both $O_2$ and $N_2$ annealed $YBa_{2-x}Sr_xCu_3O_{7-\delta}$ samples.

The above results in RE:1113 system are very surprising when compared to RE:123 system. Firstly, the former system has tetragonal structure even when the samples were oxygen treated. Whereas, the latter system are always orthorhombic after oxygen treatment. Secondly, in the RE:1113, the $T_c$ is very sensitive to the choice of RE. In comparison, for RE:123 system, $T_c$ hardly depends on the choice of RE. The answer to these nontrivial differences in the two systems can lie in the intermixing of $RE^{3+}$, $Ba^{2+}$ and $Ca^{2+}$ cations in the case of RE:1113 system [5]. Note that the crystal structure RE:1113 is derived from RE:123 and one may call RE:1113 as "tetragonal RE:123". RE:123 and RE:1113 differ only to the extent that in the latter RE, Ba and Ca are intermixed on the two usual RE:123 sites of RE and Ba in the unit cell. Thus some RE will go at Ba-site and consequently same amount of Ca will have to go to RE-site in a phase pure RE:1113 system. In fully oxygenated RE:123, all O(1) sites are full and all O(5) sites are vacant leading to formation of well known Cu-O chains characteristic of its orthorhombic structure. However, in RE:1113 system, the intermixing of RE/Ba/Ca involves aliovalent substitution of cations $RE^{3+}$ at $Ba^{2+}$ site and $Ca^{2+}$ at $RE^{3+}$ site, it disturbs the charge reservoir layer (the layer containing Cu-O chains) and oxygen rearranges in them. The rearrangement of the oxygen results in equal occupancy of O(1) and O(5) sites in RE:1113 and makes the structure tetragonal.

One is thus tempted to relate the decrease of $T_c$ in RE:1113 in comparison to RE:123 with the disorder induced in the Cu-O chains by the intermixing of RE/Ba/Ca in the former. However, the difference in the $T_c$ of La:1113 and Nd:1113 cannot be understood by different degree of intermixing in the two systems. Since the ionic size of $Ba^{2+}$ is 1.52 ?, the larger La ion (ionic size = 1.18 ?) will substitute more at Ba-site as compared to Nd ion (ionic size = 1.12 ?). This also means that equivalently more Ca goes to La-site in La:1113 as compared to Ca at Nd-site in Nd:1113. This shows that intermixing will be higher in La:1113 as compared to Nd:1113 and a higher $T_c$ is expected in the latter samples, which is in total contradiction to the obtained results. Thus we may conclude that disorder in Cu-O chains is not determining the $T_c$ in the RE:1113 system and there are other influencing factors.

One such factor related to the answer of the puzzle of different $T_c$ in La:1113 and Nd:1113 may lie in the coordination number of La and Nd. The larger $La^{3+}$ ion exists in 10 fold coordination, whereas $Nd^{3+}$ exists in 8/9 fold coordination only. Thus, when Nd substitutes Ba-site with 10 fold coordination, oxygen vacancies may get created in the adjoining $CuO_2$ layer (see also Ref.6) to satisfy the 8/9 fold coordination of Nd. Since superconductivity resides in $CuO_2$ planes, disorder in them can cause stronger degradation of $T_c$ in the case of Nd:1113. In case of La:1113, however, such disorder may not arise due to 10 fold coordination of La substituting at Ba-site.

In conclusion different $T_c$s in La:1113 ($T_c$=71 K) and Nd:1113 ($T_c$=43 K) is puzzling and the answer may lie in the formation of oxygen vacancies in $CuO_2$ planes in the latter.